\documentclass[iop]{emulateapj}
\usepackage{graphicx} 
\usepackage{times}
\usepackage{float}
\usepackage{rotating}
\usepackage{epstopdf}
\usepackage{multirow}
\usepackage{amsmath}
\usepackage{hyperref}
\usepackage{subfig}
\usepackage{url}

\shortauthors{K. De et al.}
\shorttitle{First detection of MSP microstructure}

\begin{document}
\title{Detection of polarized quasi-periodic microstructure 
emission in millisecond pulsars}
\author{Kishalay De$^1$\altaffilmark{*}, Yashwant Gupta$^2$ and Prateek 
Sharma$^1$}
\affil{$^1$Department of Physics, Indian Institute of Science, Bangalore 
560012, 
India\\
$^2$National Centre for Radio Astrophysics, TIFR, Pune University Campus, 
Post Bag 3, Pune 411007, India\\
}
\altaffiltext{*}{Present address: California Institute of 
Technology, 1200 E. California Blvd, Pasadena 91125, USA;
\href{kde@caltech.edu}{kde@caltech.edu}}

\begin{abstract}
Microstructure emission, involving short time scale, often quasi-periodic, 
intensity fluctuations in subpulse emission, is well known in normal period 
pulsars. In this letter, we present the first detections of quasi-periodic 
microstructure emission from millisecond pulsars (MSPs), from Giant Metrewave 
Radio Telescope (GMRT) observations of two MSPs at 325 and 610 MHz. Similar to 
the characteristics of microstructure observed in normal period pulsars, we 
find that these features are often highly polarized, and exhibit quasi-periodic 
behavior on top of broader subpulse emission, with periods of the order of a 
few $\mu$s. By measuring their widths and periodicities from single pulse 
intensity profiles and their autocorrelation functions, we extend the 
microstructure timescale - rotation period relationship by more than an order 
of magnitude down to rotation periods $\sim$ 5 ms, and find it to be consistent 
with the relationship derived earlier for normal pulsars. The similarity of 
behaviour is remarkable, given the significantly different physical properties 
of MSPs and normal period pulsars, and rules out several previous speculations 
about the possible different characteristics of microstructure in MSP 
radio emission. We discuss the possible reasons for the non-detection of these 
features in previous high time resolution MSP studies along with the physical 
implications of our results, both in terms of a geometric beam sweeping model 
and temporal modulation model for micropulse production.
\end{abstract}

\keywords{pulsars: general --- pulsars: individual (J0437-4715, J2145-0750) --- 
methods: observational --- techniques: polarimetric}

\section{Introduction}

Millisecond pulsars (MSPs), having rotation periods ranging from about one 
millisecond to few tens of milliseconds, have contributed significantly to our 
understanding of stellar evolution and the pulsar radio emission process. 
Believed to arise as the end products of mass accretion from a companion, the 
high rotation frequencies of these objects imply that the co-rotating 
magnetosphere (and hence the radio emission region) is significantly smaller in 
these sources compared to normal period pulsars. Nonetheless, previous studies 
have shown that although MSPs have drastically different physical parameters as 
compared to those of normal period pulsars, there are striking similarities in 
their radio emission properties as well as interesting differences (e.g.  
Kramer et al. 1998; Sallmen 1998; Kramer et al. 1999a).\\

Single pulses in normal period pulsars are known to exhibit a rich variety of 
short time-scale variations. In particular, microstructure emission has been 
long known to be a common feature of single pulse emission (first detected by 
Craft et al. 1968), and typically involves intensity variations at milli-period 
timescales, corresponding to timescales of a few hundred microseconds for 
normal period pulsars (e.g. Hankins 1972). Previous studies have shown that 
these features have typical widths for a given pulsar, and often appear as a 
train of 
quasi-periodic pulses, with their timescales showing a strong correlation with 
the pulsar period (e.g. Kramer et al. 2002; Mitra et al. 2015). \\

The characterization of microstructure emission in MSPs still remains as one of 
the elusive aspects of the field. The single pulse properties (excluding giant 
radio pulses) of only 4 MSPs have been explored till date --- PSR J0437-4715 
(Jenet et al. 1998), PSR B1534+12 (Sallmen 1998), PSR J1713+0747 (Liu et al. 
2016) and PSR B1937+21 (Jenet et al. 2001). Jenet et al. 1998 did one of the 
highest time resolution single pulse studies ever done on MSPs, observing the 
MSP J0437-4715 at L band, and reported that there was no evidence of 
microstructure down to a time resolution of 80 ns. Similar reports were 
provided by Sallmen 1998 for B1937+21 and B1534+12, and by Liu et al. 2016 for 
J1713+0747, claiming the absence of microstructure down to sub-microsecond time 
resolutions. However, the very limited number of such studies, particularly 
with high time resolution and sensitivity at low frequencies, leave a 
significant parameter space of this field unexplored. \\

Observations of sub-structures in MSP single pulses is challenging for a 
good number of reasons. On one hand, they are known to be relatively less 
luminous compared to their slower counterparts, and on the other, they require 
significantly higher time resolutions to study. While observations at low 
frequencies are favorable owing to the steep spectra of these sources, the 
effects of the interstellar medium, unless removed, pose a challenge to very 
high time resolution studies. Hence, in order to do such studies, one requires 
a telescope with very high sensitivity to be capable of detecting single 
pulses with good SNRs, as well as an appropriate backend capable of mitigating 
the dispersive effects of the interstellar medium with coherent dedispersion 
over reasonably large bandwidths. The Giant Metrewave Radio Telescope (GMRT) 
provides a unique platform to revisit some important unanswered questions on 
the 
single pulse properties of MSPs, capitalizing on its high sensitivity at metre 
wavelengths, and the recent development of a real-time coherent dedispersion 
system (De \& Gupta 2016). In this paper, we present results from a study of 
the full polarization properties of single pulses in two millisecond pulsars, 
concentrating on the characteristics of microstructure emission and its 
timescales.\\

\section{Observations}

The GMRT is a multi-element aperture synthesis telescope, consisting of 30 
antennas, each with a diameter of 45 metres, separated by a maximum baseline of 
25 km (Swarup et al. 1997). It can also be used for studies of compact objects 
like pulsars, by adding the signals from the antennas in an array mode (Gupta 
et al. 2000). We present observations of two MSPs (PSR J0437-4715 and PSR 
J2145-0750) at 325 MHz and 610 MHz, selected based on their catalogued fluxes 
at 400 MHz (S400 $>=$ 100 mJy) and Dispersion Measures (DM $<=$ 20 pc/cc). 
As we expect to see microstructure emission at time scales of 0.001 P, the 
criterion for the 
selection of frequency bands was such that the expected scattering time scale 
at each frequency band (see Table \ref{tab:obsList}) was smaller than the 
expected microstructure timescales. Wherever possible, we divided our observing 
time into multiple epochs to capitalize on serendipitous increase in the pulsar 
flux density due to refractive scintillation.\\

The observations were carried out using the Phased Array (PA) voltage beam mode 
(typically combining 14 - 16 antennas) of the GMRT Software Backend (GSB; Roy 
et al. 2010), which records the full Nyquist time resolution dual polarization 
voltage spectra for a given bandwidth, from a single phased array beam of the 
telescope. The recorded voltages were subsequently coherently dedispersed using 
an offline version of the recently developed real-time coherent dedispersion 
pipeline (De \& Gupta, 2016). The coherently dedispersed voltages (for two 
circular polarizations) were used to construct Stokes spectra with the desired 
time resolution (see Table \ref{tab:obsList}). The time resolution used was 
optimized to detect sub-structures in single pulses with enough SNR, as well as 
to have similar phase resolutions for the two pulsars. Frequency dependent 
instrumental gains were corrected using the time averaged spectra for each 
polarization signal, while frequency dependent phases due to instrumental 
delays and Faraday rotation were corrected by fitting the linear polarisation 
angle of the average profile as a function of frequency across the band. The 
specific details of the observing parameters used for each epoch, along with 
the known parameters of the pulsars are given in Table \ref{tab:obsList}. \\

\begin{table*}[!ht]
\centering
\begin{tabular}{lccccccccccc}
\hline
Pulsar Name & Frequency Band & Bandwidth & Duration & NP & Resolution & MJD & 
Period & DM & S400 & RM & $\tau_{scatt}$\\
 & MHz & MHz & sec & & $\mu$s (deg) & & ms & pc/cc & mJy & rad m$^{-2}$ & 
$\mu$s\\
\hline
J0437-4715 & 610 & 32 & 1500 & 260534 & 0.96 (0.06) & 57367 & 5.7574 & 2.6448 & 
550 & 0.0 & 0.004\\
J2145-0750 & 610 & 32 & 1920 & 119608 & 3.84 (0.08) & 57367 & 16.0524 & 8.9976 
& 100 & -1.3 & 0.031\\
J2145-0750 & 325 & 32 & 600 & 37377 & 3.84 (0.08) & 57367 & 16.0524 & 8.9976 
& 100 & -1.3 & 0.351\\
J0437-4715 & 325 & 32 & 1200 & 208427 & 0.96 (0.06) & 57380 & 5.7574 & 2.6448 & 
550 & 0.0 & 0.047\\
J2145-0750 & 325 & 32 & 1500 & 93444 & 3.84 (0.08) & 57381 & 16.0524 & 8.9976 
& 100 & -1.3 & 0.351\\
J0437-4715 & 610 & 32 & 900 & 156320 & 0.96 (0.06) & 57389 & 5.7574 & 2.6448 & 
550 & 0.0 & 0.004\\
\hline
\end{tabular}
\caption{Observing parameters for the pulsar observations carried out in this 
campaign, along with previously known parameters of the respective pulsars. 
Column 1 indicates the names of the observed sources. Column 2 
indicates the frequency band(s) where the source was observed, while Column 3 
lists the bandwidth(s) used . Columns 4 and 5 provide the duration of 
the observations, and the total number of pulses observed (NP), respectively. 
Column 6 indicates the time resolution used (in $\mu$s / degrees of phase)for 
the analysis, while Column 7 indicates the MJD of observation. Column 8, 9, 10 
and 11 list the pulsar period, DM, Flux at 400 MHz (S400) and RM respectively 
(taken from the ATNF Pulsar Catalog available at 
\url{http://www.atnf.csiro.au/people/pulsar/psrcat/}; Manchester et al. 2005). 
Column 12 indicates the expected scattering time scale for the pulsar at the 
given frequency band (derived from Cordes \& Lazio 2002).} 
\label{tab:obsList}
\end{table*}

\section{Data Analysis}
The baselines for the individual Stokes parameters $I$, $Q$, $U$ and $V$ were 
adjusted to zero mean for all the data sets. We use the symbol 
$L=\sqrt{Q^2+U^2}$ to represent the total linear polarized intensity, and 
$PA=\frac{1}{2}\arctan{\frac{U}{Q}}$ to represent the angle of the linear 
polarization. For the individual folded profiles, we plot $PA$ only if $L$ 
exceeds three times the baseline root mean square (RMS) noise. For PSR 
J2145-0750, we did not detect a significant number of single pulses with high 
SNR (peak$>$10 baseline rms) in our 325 MHz data, and hence we do not include 
this data in our subsequent analysis. The folded profiles from our observations 
at each frequency band are given in Figures \ref{fig:J0437-4715_sample} and 
\ref{fig:J2145-0750_sample}, for the pulsars J0437-4715 (at 325 
MHz and 610 MHz) and J2145-0750 (at 610 MHz) respectively, which are consistent 
with previously published polarization profiles at these frequencies (Navarro 
et al. 1997; Stairs et al. 1999). \\

\subsection{Autocorrelation Function}
The timescales of microstructure emission, i.e., their widths and periodicity 
(if present) are of fundamental importance in microstructure studies. The 
Autocorrelation Function (ACF) is a useful tool to measure these timescales in 
individual pulses, as well as to detect the presence of `preferred' timescales 
(if any) in a large ensemble of pulses. We use the definition of the ACF given 
in Lange et al. 1998 for our analysis. In normal period pulsars, the ACFs of 
individual subpulses are primarily dominated by a smooth bell shaped curve 
centered at zero lag, with micropulses forming weak (or occasionally strong) 
modulations on the broader decline of the ACF. In cases where quasi-periodic 
microstructure is present, they form periodic maxima in the ACF, with the time 
lag of the first peak representing the characteristic separations of the 
micropulses (e.g. Cordes at al. 1990).\\

\subsection{Searching for periodic microstructure}
We searched for the presence of microstructure in our data set by browsing 
through hundreds of single pulses from each observation. While we did attempt 
to find the presence of microstructure numerically, as done previously using 
ACF slope-flattening / maxima searches (e.g. Lange et al. 1998), we found that 
the diverse morphology of single pulses, along with the lack of a large number 
of single pulses with very high SNR made such automated techniques difficult to 
implement accurately. Hence, for each pulsar (at each frequency band), 
we manually browsed through the brightest single pulses (with a peak SNR 
threshold above a certain value) closely examining their morphology and ACFs. 
When periodic micropulses were found by visual inspection of the single pulse 
time series, we cross-checked the ACF for the presence of periodic maxima 
consistent with the separations and widths of the time domain features, and if 
present, recorded the first maxima in the ACF as the periodicity of the 
micropulses. Since both pulsars in our sample have low DMs, we also examined 
the off-pulse region around pulses which show microstructure, along with the 
corresponding zero DM time series, to exclude the possibility of RFI features 
mimicking microstructure, and did not find evidence for such contamination. 
Unless mentioned otherwise, we refer only to timescales of Stokes I in 
subsequent discussion.\\

\section{Results}

\subsection{PSR J0437-4715}
\begin{figure*}[!ht]
\centering
\raggedleft
\subfloat[]{\includegraphics[width=\columnwidth]{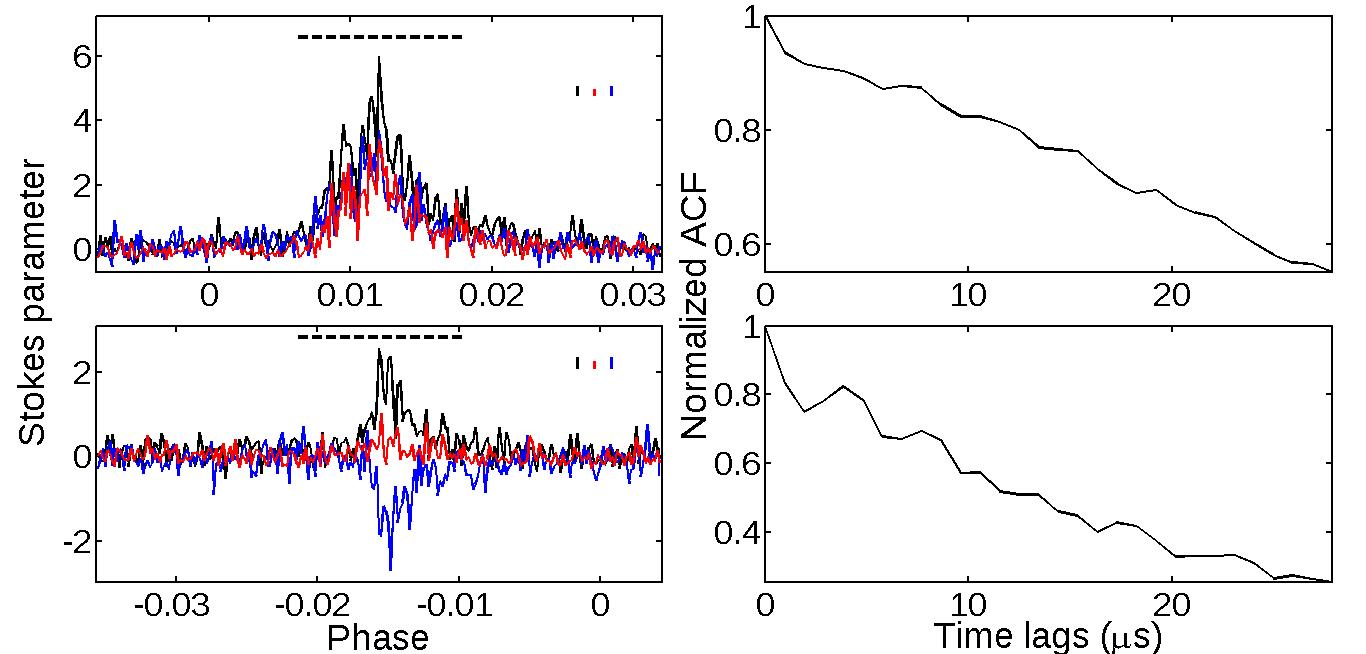}}
\raggedright
\subfloat[]{\includegraphics[width=\columnwidth]{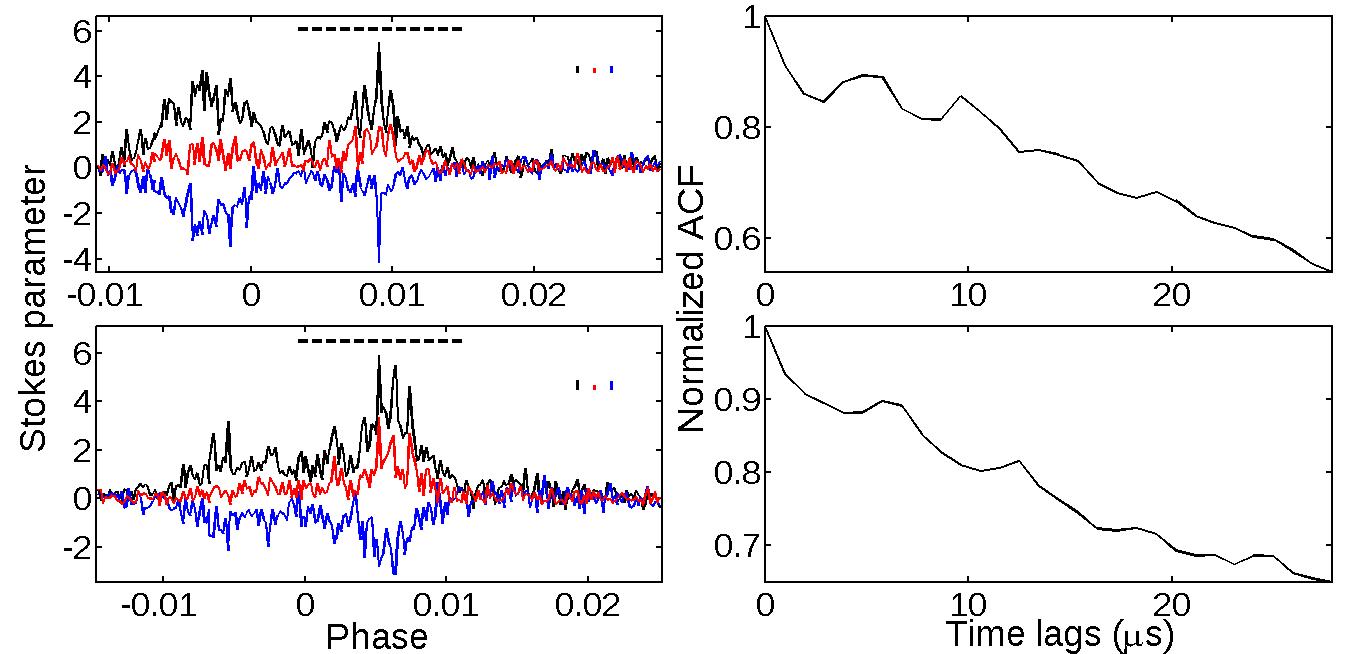}}
\\
\raggedleft
\subfloat[]{\includegraphics[width=\columnwidth]{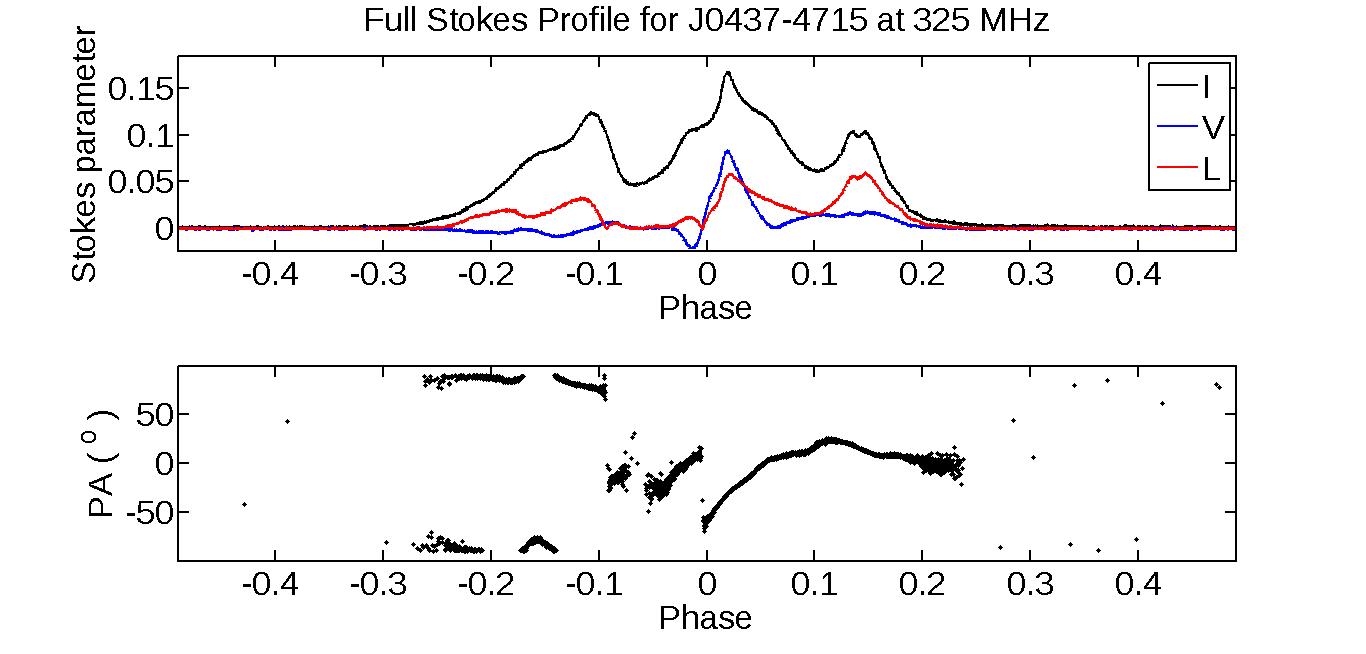}}
\raggedright
\subfloat[]{\includegraphics[width=\columnwidth]{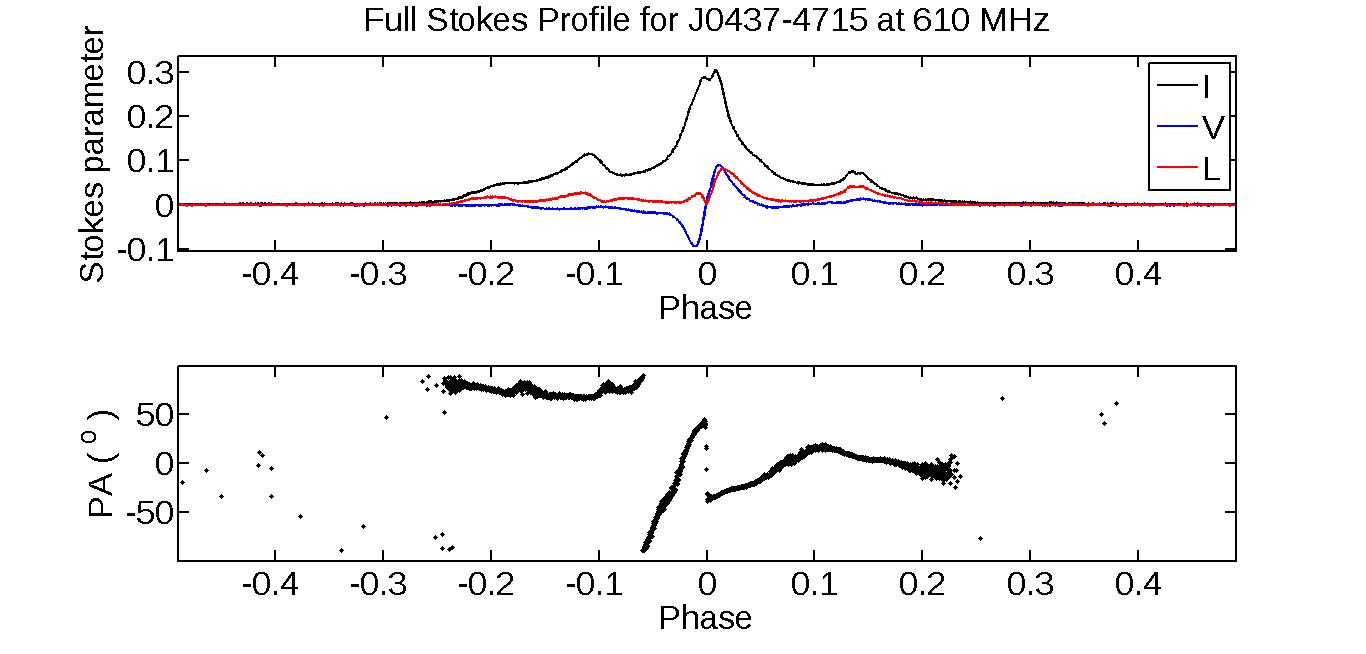}}	
 
\caption{(a) Left panels show Stokes I (black), L (red), V (blue) plots of two 
samples pulses from PSR J0437-4715 at 325 MHz exhibiting quasi-periodic 
microstructure, along with their respective Stokes I ACFs on the right. The 
ACFs have been computed for the subpulse region under the dashed line. 
The colored bars on the top right of the panels containing the sample 
pulses indicate the off-pulse RMS of the respective Stokes parameters. (b) Same 
as (a) for PSR J0437-4715 at 610 MHz. (c) Folded polarization profile of 
J0437-4715 from an observation at 325 MHz. (d) Folded polarization profile of 
J0437-4715 at 610 MHz. Note the prominent oscillations in the single pulse ACFs 
produced by quasi-periodic microstructure.}
\label{fig:J0437-4715_sample}
\end{figure*}

We found a large number of single pulses with high SNR (with threshold set at 
peak $>$ 15 baseline RMS) from this source --- detecting $>$ 100 pulses at 325 
MHz and $>$ 800 pulses at 610 MHz . The brightest single pulses are 
often highly polarized narrow spikes of emission, with widths $<$ 10 $\mu$s and
no sub-structure apparent down to our time resolutions, consistent with 
previous reports by Jenet et al. 1998. Manually browsing these pulses, we 
found a significant number ($\sim$ 10 \%) of pulses exhibiting microstructure 
at 610 MHz (which are not necessarily the brightest pulses in the data set). We 
further detected microstructure in a few bright pulses at 325 MHz, although the 
number of such pulses was lower due to the relatively lower SNR of the data. A 
few such sample pulses from this source, exhibiting polarized quasi-periodic 
microstructure are shown in Figure \ref{fig:J0437-4715_sample}. \\

Micropulses, when present, are typically highly polarized, and appear either 
in the form of a train of 3 or more quasi-periodic modulations, or in the form 
of single pairs of micropulses. The modulations appear in Stokes I, L and V, 
and are typically weak compared to the underlying subpulse emission. As 
most micropulses appear in the form of close unresolved trains, it was not 
possible to measure their widths from the single pulse ACFs. Hence, we 
recorded only the micropulse periodicities from the ACFs, and show their 
distribution in Figure \ref{fig:periodDist}, along with the SNR distribution of 
all pulses and pulses which exhibit microstructure. We note, however, that 
when micropulses are resolved, their widths are typically $\sim$ 2 $\mu$s, and 
hence unresolved with typical periodicities of $\sim$ 4 $\mu$s. The smaller 
number of bright pulses exhibiting microstructure at 325 MHz did not allow us 
to construct meaningful distributions. Nonetheless, we report a median 
periodicity of 4.72 $\mu$s found in quasi-periodic microstructure for this 
source at 325 MHz.\\

\subsection{PSR J2145-0750}
\begin{figure*}[!ht]
\centering
\raggedleft
\subfloat[]{\includegraphics[width=\columnwidth]{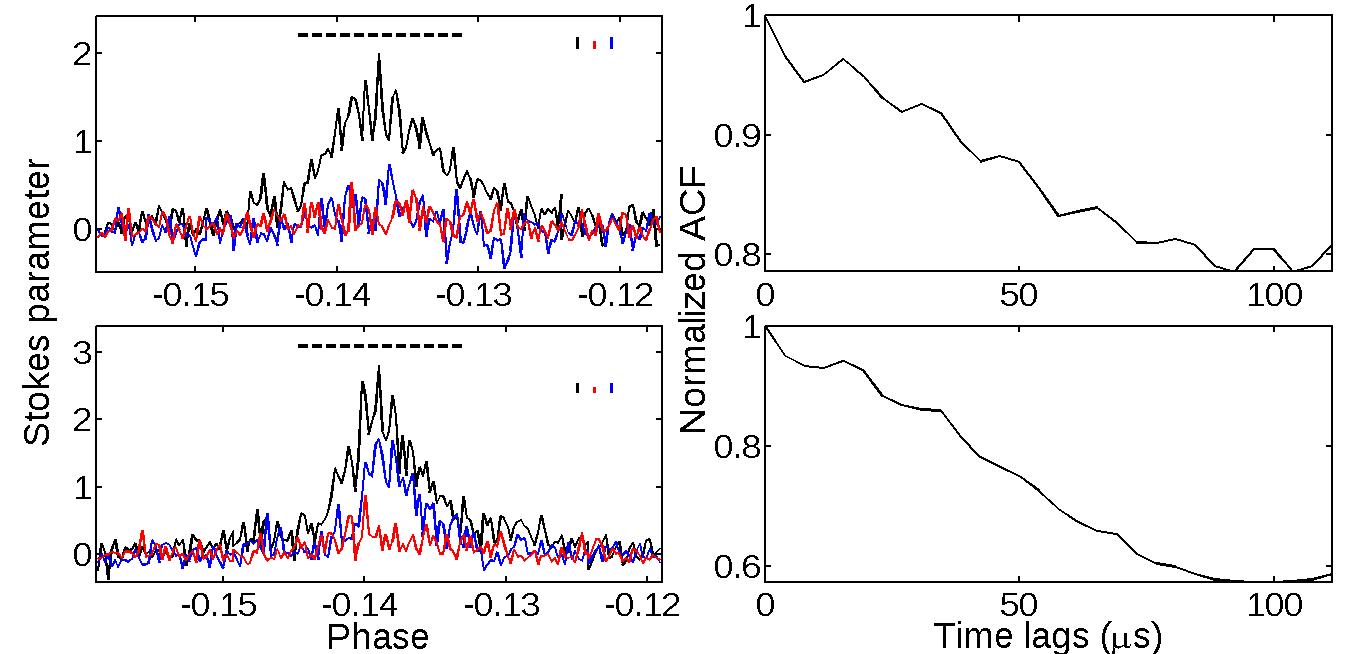}}
\raggedright
\subfloat[]{\includegraphics[width=\columnwidth]{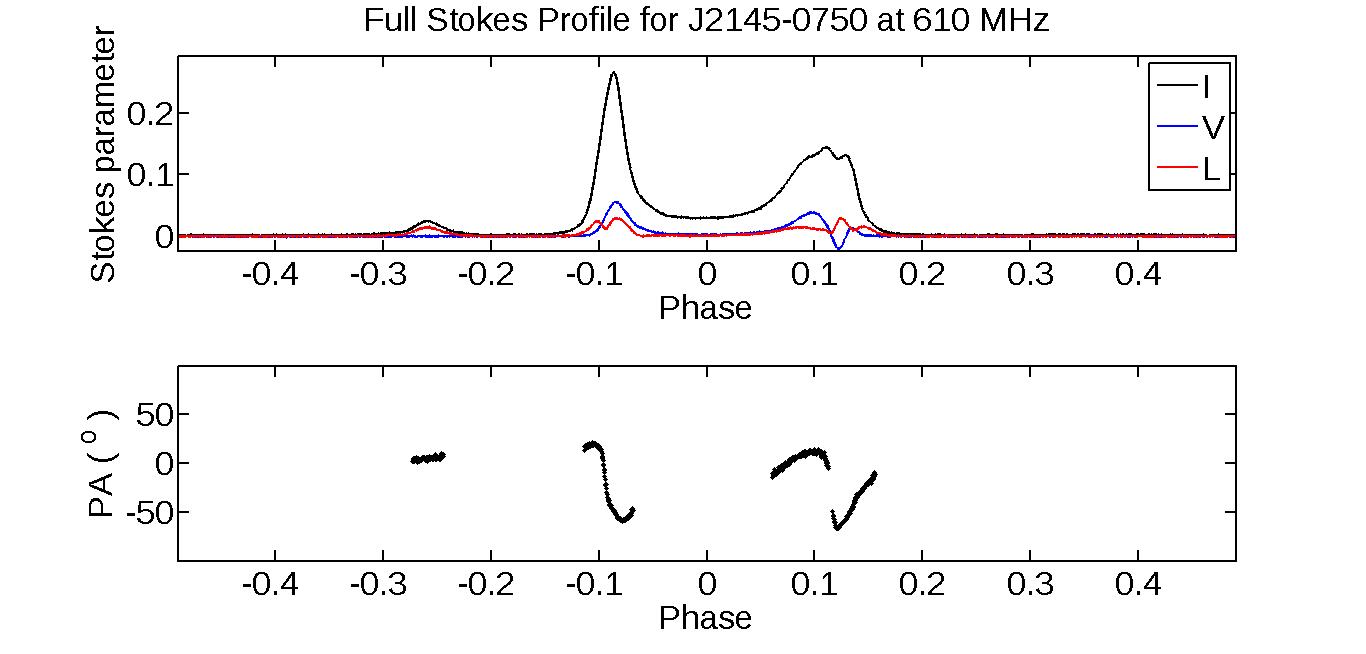}}
\caption{(a) Same as Figure \ref{fig:J0437-4715_sample} (a) for PSR J2145-0750 
at 610 MHz. (b) Folded polarization profile of J2145-0750 from an observation 
at 610 MHz. Note the prominent periodic oscillations in the single pulse ACFs 
produced by periodic microstructure.}
\label{fig:J2145-0750_sample}
\end{figure*}

We did not detect a significant number of single pulses with high SNR (peak $>$ 
10 baseline RMS) in any of the two observing epochs for this source at 325 MHz. 
However, in the single observing epoch at 610 MHz, we found $>700$ single 
pulses with peak $>$ 15 baseline RMS. Browsing these single pulses, we found a 
good fraction ($\sim$ 15 \%) of the pulses to exhibit polarized quasi-periodic 
microstructure. Similar to J0437-4715, the brightest pulses are devoid 
of microstructure and exhibit a broad emission component only, while the 
relatively fainter pulses exhibiting microstructure consist of a broad subpulse 
component with weak modulations. A few typical pulses exhibiting 
quasi-periodic microstructure are shown in Figure \ref{fig:J2145-0750_sample}.\\

\begin{figure*}[!ht]
\centering
\raggedleft
\subfloat[]{\includegraphics[width=\columnwidth]{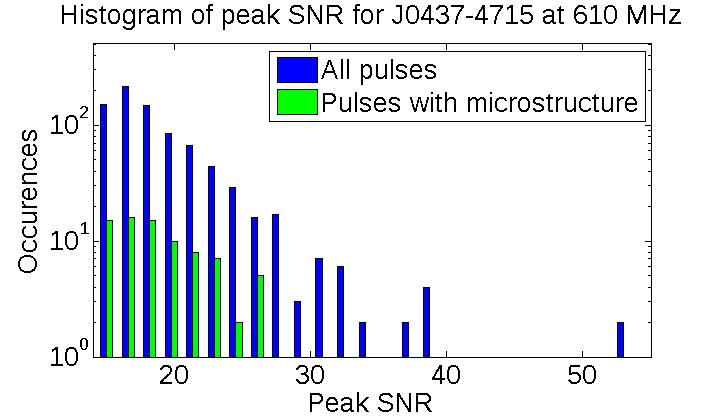}}
\raggedright
\subfloat[]{\includegraphics[width=\columnwidth]{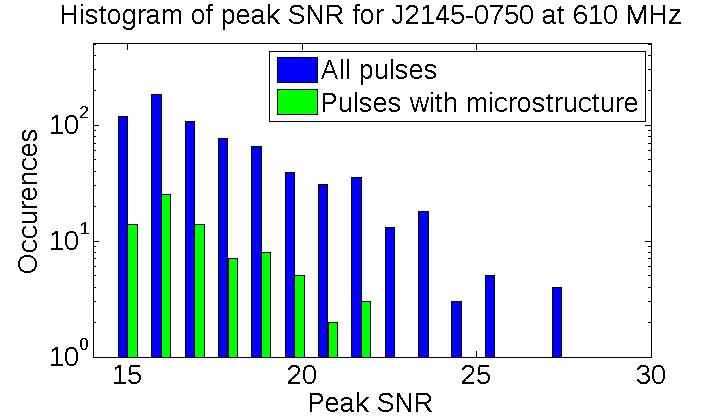}}
\\
\raggedleft
\subfloat[]{\includegraphics[width=\columnwidth]{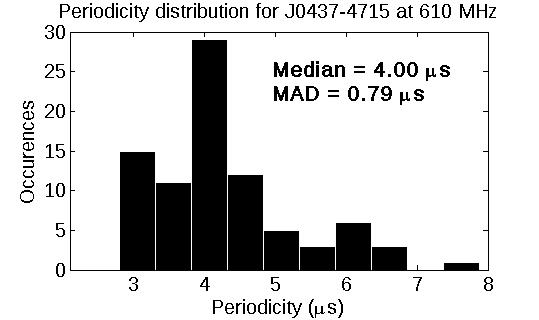}}
\raggedright
\subfloat[]{\includegraphics[width=\columnwidth]{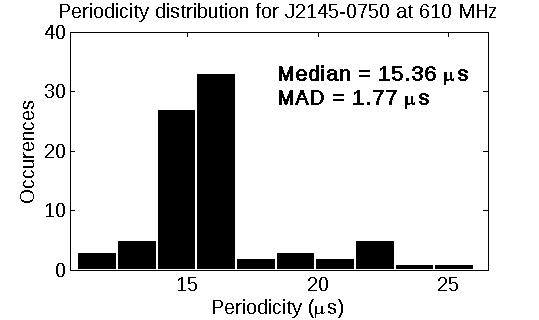}}	
\caption{Distributions of peak SNR for all pulses (blue) and pulses 
showing microstructure (green) are shown for J0437-4715 (a) and J2145-0750 (b). 
Note that although the two distributions appear to scale with each other 
closely at lower SNR, the brightest pulses do not show microstructure. Also 
shown are distributions of microstructure quasi-periodicities for J0437-4715 
(c) and J2145-0750 (d) observed at 610 MHz. The median periodicity and median 
absoulte deviation (MAD) of the distributions is indicated in the figure. Note 
that these distributions have a prominent maximum indicating the presence of a 
preferred timescale, with a Kolmogorov Smirnov null hypothesis (of the 
distribution having no preferred timescale, i.e., being a uniform distribution) 
probability of $\sim$ 10$^{-7}$ in both cases.}
\label{fig:periodDist}
\end{figure*}

We find micropulses to be highly polarized in this source, occasionally showing 
very high levels of circular polarization (see for example, the lower panel in 
Figure \ref{fig:J2145-0750_sample} (a)). Resolved micropulses are virtually 
absent in this source, as they always appear in quasi-periodic trains of 4 or 
more micropulses. As a result, we recorded only the periodicities of 
the microstructure from the single pulse ACFs, and show their distribution in 
Figure \ref{fig:periodDist}, along with the distribution of SNR, as in the case 
of J0437-4715. Nonetheless, the microstructure periodicities suggest that the 
individual micropulse widths are $\gtrsim$ 8 $\mu$s.\\

\section{Discussion}
The origin of microstructure emission remains as one of the most perplexing 
unanswered questions in our understanding of pulsar radio emission even though 
it is well established that microstructure is indeed an integral component of 
the emission process. In this paper, we have reported on the first detections 
of microstructure emission in millisecond pulsars, which suggests that despite 
of the significant differences in the physical parameters of their 
magnetospheres, microstructure remains as one of the fundamental properties of 
radio emission in these sources. Indeed, the observation that the pulses 
exhibiting microstructure are typically \textit{not} the brightest pulses in 
our data set suggests that there is likely an ensemble of weaker pulses below 
our detection threshold which exhibit microstructure, thereby indicating that 
it may be a fairly common occurrence in MSPs.  \\

It is instructive to reflect upon the reasons for non-detection of these 
features in previous high time resolution studies of MSPs. Firstly, it is 
important to note that the majority of previous single pulse studies of MSPs 
have had very limited sensitivity towards detection of single pulses, and have 
concentrated on statistics of accumulated pulses over many periods (e.g. Jenet 
et al. 2001). Our study is one of the first reports of high sensitivity, high 
time resolution observations at relatively low frequencies (at and below 600 
MHz), which forms an ideal combination for detection of micropulses from MSPs. 
Interestingly, Jenet et al. 1998 did a study on J0437-4715 with similar 
sensitivities at 1.4 GHz, and claimed that there was no evidence for 
microstructure emission. However, we note that their analysis concentrated on a 
search for microstructure features in the average ACF of single pulses, which 
often does not show the expected `breaks' when microstructure is weak or does 
not have a preferred timescale (Cordes et al. 1990). More recently, Liu et al. 
2016 reported on the non-detection of microstructure emission in PSR J1713+0747 
at L band, by analyzing the properties of the averaged ACF as well as single 
pulse ACFs of the brightest pulses. We note that higher frequency 
non-detections may also be due to a steeper spectrum of micropulses compared to 
subpulse emission (as shown for B2016+28 by Cordes et al. 1990).\\

\begin{figure}[!ht]
\centering
\figurenum{4}	
\plotone{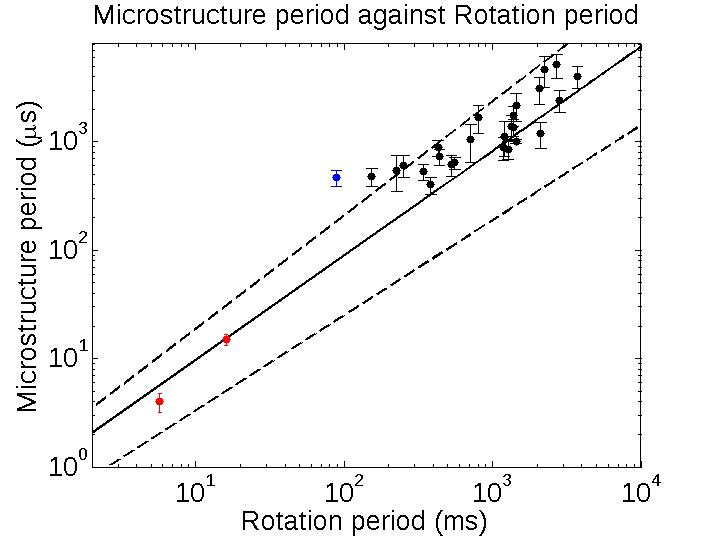}
\caption{Microstructure periods of pulsars with known microstructure properties 
plotted against their rotation period. The red points are the ones added from 
this study, while the blue point corresponds to the reported microstructure 
periodicity for the Vela pulsar --- the fastest rotating pulsar for which 
microstructure had been studied previously. The solid line corresponds to the 
best fit power law relationship between these quantities, while the dashed 
lines correspond to the 95\% confidence interval on this relationship.}
\label{fig:microRotRel}
\end{figure}

We now use our results to examine the microstructure quasiperiodicity - pulsar 
rotation period relationship down to the millisecond rotation periods, using 
the median quasi-periodicity of the distributions (Figure \ref{fig:periodDist}) 
as the typical periodicity and their median absolute deviation (MAD) as the 
characteristic scatter. Since we find micropulses to be mostly unresolved from 
the underlying quasi-periodicity, it is difficult to estimate their intrinsic 
widths and hence, we do not discuss the micropulse width - rotation period 
relationship. We use previously reported microstructure periodicites in Kramer 
et al. 2002 (Table 2) and Mitra et al. 2015 (Table 3) along with our new 
results to extend the relationship. Note that previous reports combine 
observations from multiple frequencies of observation, and in cases where 
multiple reports were present (at different frequencies, different analysis 
techniques or for different subpulse components), we only used the results with 
reported uncertainty estimates and averaged their timescales for our analysis. 
While there is some weak evidence for change in microstructure timescales 
across frequencies, the dependence is known to be very weak (e.g. Kramer et al. 
2002; Hankins et al. 1976; Ferguson \& Seiradakis 1978). Hence, we do not 
expect such averaging to change the properties of the relationship 
significantly. As shown in Figure \ref{fig:microRotRel}, we derive a 
relationship of 
\begin{equation}
 P_{\mu} = (1.06 \pm 0.62) P_r^{(0.96 \pm 0.09)}
\end{equation}
where $P_{\mu}$ is the typical microstructure periodicity in units of $\mu$s 
and $P_r$ is the rotation period in units of ms. The uncertainties shown are at 
95\% confidence level. The relationship is consistent with the linear 
relationships derived by Cordes et al. 1979 and Mitra et al. 2015 from data for 
only normal period pulsars. This result indicates that there is a striking 
similarity of the properties of microstructure emission in normal and recycled 
pulsars.\\

Lastly, we note that while we have presented here an analysis of the timescales 
of Stokes I only, we were also able to measure the periodicities of Stokes V 
and Stokes L in cases where micropulses were highly polarized. While we did 
find that the periodicities in Stokes I, V and L were largely similar, the 
lower SNR of the Stokes V and L profiles prevented a statistically accurate 
conclusion about their consistency. Additionally, the low SNR of the Stokes L 
profiles did not allow a close examination of the behavior of the PA across 
individual micropulses. Indeed, recent reports by Mitra et al. 2015 have 
suggested that the periodicities in these parameters are consistent with each 
other, leading them to argue that microstructure \textit{does not} arise from 
single particle curvature radiation in vacuum. \\

However, the stability of the micropulse timescale - rotation period 
relationship down to millisecond periods does support a geometric origin of 
microstructure, wherein micropulses are formed by sweeping beams of radiation 
arranged longitudinally across the emission region (with constant angular 
widths $\sim$ 0.18$^o$). We note that the reported inconsistency of the 
polarization signatures only rule out single particle curvature radiation as 
the 
origin of micropulse emission, but do not necessarily rule out a geometric 
origin. Additionally, a radial / temporal origin of micropulse production is 
also a plausible alternative. In this model, micropulses arise from a radial 
modulation of the emission region, created either during the plasma production 
process or by modulation during propagation. It must be noted, however, that 
the 
modulation mechanism \textit{must} be directly associated with the pulsar 
rotation period (for example, by means of rotation dependent electromagnetic 
fields; Cordes et al. 1990) to account for the strong dependence of periodicity 
with the rotation period. Clearly, future theoretical progress in this field 
coupled with observations of a larger number of pulsars are required to resolve 
these issues.\\

\section{Summary}
In this paper, we have presented the first report of detections and statistics 
of quasi-periodic microstructure emission in recycled millisecond pulsars. We 
find these micropulses to be often highly polarized, forming typically weak 
modulations on a broader subpulse envelope. By estimating microstructure 
periodicities from the single pulse ACFs, we report the first detections of 
`preferred' microstructure periodicity in millisecond pulsars. This result 
allows us to extend the microstructure timescale - rotation period relationship 
by more than an order of magnitude, and thus put constraints on the geomteric 
sweeping and temporal modulation models of micropulse emission. While we cannot 
draw statistically significant conclusions about the detailed polarization 
signatures of micropulses from our data, future high sensitivity observations 
at low frequencies are expected to resolve these aspects. In particular, the 
significantly larger bandwidths of the upgraded GMRT (uGMRT), which is 
currently being commissioned, is likely to allow an even larger scale study of 
microstructure emission in millisecond pulsars, encompassing not only their 
full polarization properties but also their spectral evolution characteristics, 
explored using simultaneous dual-frequency observations.\\

\section*{Acknowledgements}
We thank the staff of the GMRT who have made these observations possible. GMRT 
is run by the National Centre for Radio Astrophysics of the Tata Institute of 
Fundamental Research. We would like to thank R. T. Gangadhara for valuable 
discussions, as well as the anonymous referee for the comments that 
helped significantly improve the paper. KD acknowledges support from a 
fellowship provided by the Kishore Vaigyanik Protsahan Yojana (KVPY) scheme of 
the Department of Science and Technology, Government of India. Partial support 
was provided by JC Bose Fellowship of Prof. Arnab Rai Choudhuri. This research 
has made use of NASA's Astrophysics Data System Bibliographic Services.

\end{document}